\documentclass[12pt]{article}

\newcommand{\EQ}{\begin{equation}}
\newcommand{\EN}{\end{equation}}
\newcommand{\bea}{\begin{eqnarray}}
\newcommand{\ena}{\end{eqnarray}}
\newcommand{\bdis}{\begin{displaymath}}
\newcommand{\edis}{\end{displaymath}}
\newcommand{\vs}[1]{\vspace{#1 mm}}

\renewcommand{\a}{\alpha}

\renewcommand{\d}{\delta}
\renewcommand{\v}{\Delta}
\renewcommand{\t}{\tau}

\usepackage[dvips]{graphicx}
\begin{document}

\topmargin 0pt
\oddsidemargin - 3mm

\newcommand{\NP}[1]{Nucl.\ Phys.\ {\bf #1}}
\newcommand{\PL}[1]{Phys.\ Lett.\ {\bf #1}}
\newcommand{\CMP}[1]{Comm.\ Math.\ Phys.\ {\bf #1}}
\newcommand{\PR}[1]{Phys.\ Rev.\ {\bf #1}}
\newcommand{\PRL}[1]{Phys.\ Rev.\ Lett.\ {\bf #1}}
\newcommand{\PREP}[1]{Phys.\ Rep.\ {\bf #1}}
\newcommand{\PTP}[1]{Prog.\ Theor.\ Phys.\ {\bf #1}}
\newcommand{\PTPS}[1]{Prog.\ Theor.\ Phys.\ Suppl.\ {\bf #1}}
\newcommand{\NC}[1]{Nuovo.\ Cim.\ {\bf #1}}
\newcommand{\JPSJ}[1]{J.\ Phys.\ Soc.\ Japan\ {\bf #1}}
\newcommand{\MPL}[1]{Mod.\ Phys.\ Lett.\ {\bf #1}}
\newcommand{\IJMP}[1]{Int.\ Jour.\ Mod.\ Phys.\ {\bf #1}}
\newcommand{\AP}[1]{Ann.\ Phys.\ {\bf #1}}
\newcommand{\RMP}[1]{Rev.\ Mod.\ Phys.\ {\bf #1}}
\newcommand{\PMI}[1]{Publ.\ Math.\ IHES\ {\bf #1}}
\newcommand{\JETP}[1]{Sov.\ Phys.\ J.E.T.P.\ {\bf #1}}
\newcommand{\TOP}[1]{Topology\ {\bf #1}}
\newcommand{\AM}[1]{Ann.\ Math.\ {\bf #1}}
\newcommand{\LMP}[1]{Lett.\ Math.\ Phys.\ {\bf #1}}
\newcommand{\CRASP}[1]{C.R.\ Acad.\ Sci.\ Paris\ {\bf #1}}
\newcommand{\JDG}[1]{J.\ Diff.\ Geom.\ {\bf #1}}
\newcommand{\JSP}[1]{J.\ Stat.\ Phys.\ {\bf #1}}
\newcommand{\JHEP}[1]{JHEP\ {\bf #1}}

\begin{titlepage}
\setcounter{page}{0}
\begin{flushright}
\end{flushright}

\vs{5}
\begin{center}
{\Large  Stochastic Quantization Approach for Causal Dynamical Triangulation String Field Theory }

\vs{10}

{\large Hiroshi\ Kawabe\footnote{e-mail address:
kawabe@yonago-k.ac.jp}} \\
{\em
Yonago National College of Technology \\ 
Yonago 683-8502, Japan} \\
\end{center}

\vs{8}
\centerline{{\bf{Abstract}}}

\
We construct a 2-dimensional Causal Dynamical Triangulation (CDT) model from a matrix model which represents the loop gas model of closed string.
The target-space index is reinterpreted as time or geodesic distance.
We apply stochastic quantization method to the model to obtain the Generalized CDT (GCDT), which has additional interaction of creating baby universe.
If we take a specific scaling in continuum limit, we realize an extended GCDT model characterized by the non-critical string field theory.

\end{titlepage}
\newpage
\renewcommand{\thefootnote}{\arabic{footnote}}
\setcounter{footnote}{0}

More than a decade ago, matrix model was expected to realize the non-perturbative definition of string field theories through the double scaling limit\cite{DS}.
In the matrix models, the interaction of the spin cluster domain wall (I-K type interaction) plays an important role to obtain the non-critical string field theory\cite{IK}.
Hermitian matrix models formulate the Dynamical Triangulation of the discrete 2D surface, in which orientable strings propagate or interact\cite{JR}\cite{AJ}\cite{Mog}. 
The Loop gas model is a description of the non-critical string field theory, in which every string is located in 1-dimensional discrete target-space point $x$ and interacts with another one in the same point or neighboring points in each time evolution\cite{KK}.
It is formulated by the matrix possessing the additional index $x$\cite{Kos}\cite{EKN}.
On the other hand, it is well known that the stochastic quantization of the matrix models is effective to deduce the string field theories, and the stochastic time plays the role of the geodesic distance on the 2D random surfaces\cite{JR}\cite{Nak}.
However, one of the problems in the Dynamical Triangulation is that the probability of splitting interaction becomes too large to construct the realistic space-time.
This problem becomes more severe in higher dimension.
Even in 2D model, the whole surface of the world-sheet is covered with many projections of infinitesimal baby universe.

The Causal Dynamical Triangulation (CDT) model is proposed to improve the above problem\cite{AL}.
In this model, the triangulation is severely restricted because of the time-foliation structure.
The most characteristic feature of the CDT model is that the causality forbid the splitting and merging interaction.
There appears no baby universe and a string propagator becomes a torus with smooth surface.
The CDT model is generalized to include only the splitting interaction but not the merging interaction.
It is the Generalized Causal Dynamical Triangulation (GCDT) model and baby universes make the world surface not be smooth\cite{ALWZ}.
A string field theory is constructed from the CDT and its Schwinger-Dyson (S-D) equations are investigated\cite{ALWWZ1}.
It is also formulated by a matrix model further\cite{ALWWZ2}.
Recently, the GCDT model is extended to include additional I-K type interaction and its matrix model formulation is also proposed\cite{FSW}.
The S-D equation of this model has features of the non-critical string field theory.

In this note, we construct the CDT model from the matrix model for the loop gas model.
We assign the matrix an additional discrete index, which is interpreted as discrete time or geodesic distance.
In the original loop gas model, it is interpreted as space.
Then, we apply the stochastic quantization method to this model in order to realize a string field theory of the GCDT model.
The main difference from other matrix model formulation is that the stochastic time does not have relation to the geodesic distance in our model.
\begin{figure}[t]
\begin{center}
\includegraphics [width=100mm, height=20mm]{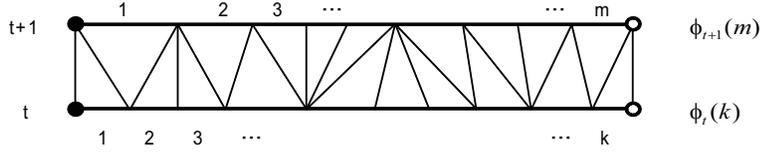}
\caption{A configulation of the CDT triangulation of the 1-step propagation}
\label{fig:CDT1}
\end{center}
\end{figure}

At the beginning, we briefly review some fundamental nature about the CDT in the 2D space-time.
In this model, a torus of loop propagation is sliced to many rings with small width $a$, the minimal discrete time.
Each ring corresponds to the 1-step time propagation of a loop, from an edge to another edge.
Loops of the edges are composed of the links with length $a$ so that the length of loop is also discretized.
A ring is constructed with triangles, one of whose three edges has to be a component of the loop on one side and the other two edges are connected to other triangles.
Therefore the 1-step propagation from a loop with the length $k$ to another one with the length $m$ is composed of $k+m$ triangles, with $k$ upward ones and $m$ downward ones (Fig.\ref{fig:CDT1}).
Since the number of the configuration corresponds to the amplitude of this 1-step propagation, we define the 1-step "two-loop function" as
\bea
\label{eq:amplitude}
G^{(0)} (k,m; 1) \equiv {g^{k+m} \over k+m}~_{k+m}{\rm C}_k ,
\ena
where $g$ is associated with each triangle and $_{k+m}{\rm C}_k$ expresses the binomial coefficient.
It is rather convenient to define the 1-step marked propagator of the loop with length $k$ as $G^{(1)} (k,m;1) \equiv k G^{(0)}(k,m;1)$.
We can construct $t$-step unmarked (and marked) propagators by piling up the 1-step unmarked (and marked) propagators
\bea
\label{eq:tstep}
G^{(0)} (n,m;t) & = & \sum_{k=1}^{\infty} G^{(0)}(n,k;t-1) k G^{(0)} (k,m;1), \nonumber \\
G^{(1)} (n,m;t) & = & \sum_{k=1}^{\infty} G^{(1)}(n,k;t-1) G^{(1)} (k,m;1),
\ena
respectively.
The disc amplitude $W(n)$ is the summation of the amplitudes such that the loop with the length $n$ becomes to zero in some future time, and it is expressed as
\bea
\label{eq:discamp}
W(n) \equiv \sum_{t}^{\infty} G^{(1)} (n,0;t).
\ena
Then, we expect the superposing relation,
\bea
\label{eq:discsum}
W(n) \equiv \sum_{k=1}^{\infty} G^{(1)} (n,k;1) W(k)+G^{(1)}(n,0;1).
\ena

Originally, the CDT model does not contain splitting interaction nor merging interaction, because these processes violate the causality.
This means that a saddle point on the world-sheet causes to two distinct light cones.
However, we can include the splitting interaction if we impose the condition such that any separated baby loop shrinks to length zero and the mother loop propagates without interacting with it.
It is the GCDT model, in which the "causality" in a broad sense is respected.
 
We propose a matrix model of the modified version of the loop gas model, with a fundamental matrix $(M_{tt'})_{ij}$, where the indices $i,~j$ run from 1 to $N$.
The $N \times N$ matrix $M_{tt'}$ corresponds to a link variable which connects two sites on the discrete times $t$ and $t'$ with the direction from $t$ to $t'$.
Then we start with the action of $U(N)$ gauge invariant form,  
\bea
\label{eq:actionM}
S[M] = -g\sqrt{N} {\rm tr} \sum_t M_{tt'} + {1 \over 2}{\rm tr} \sum_{t,t'} M_{tt'} M_{t't} - {g \over 3\sqrt{N}} {\rm tr} \sum_{t,t',t''} M_{tt'} M_{t't''} M_{t''t},
\ena
with the partition function $Z = \int {\cal D} M e^{-S[M]}$.
$M_{tt} \equiv A_t$ is a hermitian matrix, which corresponds to a link of discrete string element soaked in one time $t$.
$M_{t,t+1} \equiv B_t$ and $M_{t+1,t} \equiv B^{\dagger}_t$ are associated with a link connecting sites on the nearest neighboring times $t$ and $t+1$.
Otherwise $M_{tt'} = 0$ (for $t' \neq t, t \pm 1$).
Hence we can rewrite the partition function as $Z = \int {\cal D} A {\cal D} B {\cal D} B^{\dagger} e^{-S[A,B,B^{\dagger} ] } $ with the action 
\bea
\label{eq:actionAB}
S[A, B, B^{\dagger}] &=& -g\sqrt{N} {\rm tr} \sum_t A_{t} + {1 \over 2}{\rm tr} \sum_t A_t^2 + {\rm tr} \sum_t B_t B_t^{\dagger}  \nonumber \\
 & & - {g \over 3\sqrt{N}} {\rm tr} \sum_t A_t^3 - {g \over \sqrt{N}} {\rm tr} \sum_t (A_t B_t B_t^{\dagger} + A_{t+1} B_t^{\dagger} B_t).
\ena
\begin{figure}[t]
\begin{center}
\includegraphics [width=100mm, height=35mm]{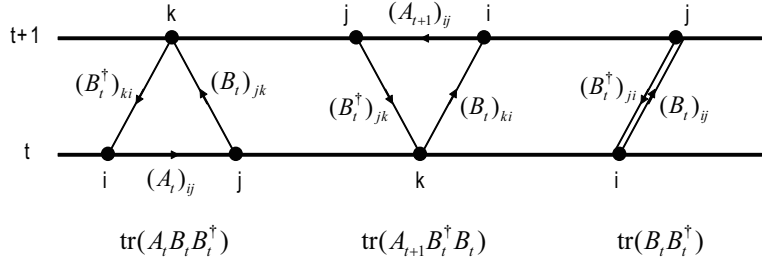}
\caption{The triangulation assignment for the cubic terms and the square term}
\label{fig:CDT2}
\end{center}
\end{figure}
Two square terms in the first line play the role of gluing the links of two triangles, and three cubic terms in the second line express the triangles which are the elements of surface (Fig.\ref{fig:CDT2}).
While the last two terms composed of $A, B$ and $B^{\dagger}$ correspond to the triangles on the surfaces of string propagation ring, the cubic term of $A$ produces a triangle soaked in one time-slice, which does not exist in the CDT model.
Through integrating out the non-hermitian matrices $B_t$ and $B^{\dagger}_t$, we obtain the effective theory only with equi-temporal link matrices $A_t$.
The partition function becomes $Z \equiv \int {\cal D} A e^{-S_{\rm eff} [A]}$, where
\begin{eqnarray*}
S_{\rm eff} [A] = {\rm tr} \sum_t \left[ -g \sqrt{N} A_t + {1 \over 2} A_t^2  - {g \over 3\sqrt{N}} A_t ^3 + {\rm log} \Bigl\{ {\bf 1} - {g \over \sqrt{N}}  \left( A_t {\bf 1}_{t+1} + {\bf 1}_t A_{t+1}  \right) \Bigr\} \right].
\end{eqnarray*}
If we define a loop variable $\phi _t (n) \equiv {1 \over N}{\rm tr}({A_t \over \sqrt{N}})^n$ as the discrete closed string of length $n$ in time $t$, the effective action of the loop variables is written as
\bea
\label{eq:actioneff}
S_{\rm eff} [\phi , A] & = & - N^2 \sum_t \left[ {g \over N} {\rm tr}  {A_t \over \sqrt{N}} - {1 \over 2N}{\rm tr}  \left( {A_t \over \sqrt{N}} \right) ^2  + { g \over 3N} {\rm tr}  \left( {A_t \over \sqrt{N}} \right) ^3 \right. \nonumber \\
 & &  \left. +  \sum_{k=0}^{\infty} \sum_{m=0}^{\infty} G^{(0)}(k,m;1) \phi _t (k) \phi _{t+1}(m) \right],
\ena
where $G^{(0)}(k,m;1)$ is the two-loop function of the 1-step time appeared in the CDT model.
The last term of eq.(\ref{eq:actioneff}) is expressed graphically in Fig.\ref{fig:CDT1}.
We may construct the $t$-step propagator as 
\begin{eqnarray*}
nmG^{(0)}(n,m;t)  = \langle \phi_0 (n) \phi _t (m) \rangle  = {1 \over Z} \int {\cal D} A \phi _0 (n) \phi _t (m) e^{-S_{\rm eff}[A]}, 
\end{eqnarray*}
This expresses the sum over all ways of connecting $\phi _0 (n)$ with $\phi _t (m)$ by $t$ times piling of 1-step two-loop functions.
Thus we realize the matrix model formulation of the CDT model.

Now, we extend this model to the GCDT with loop interactions by applying the stochastic quantization method to the above model.
The Langevin equation for a matrix variable and white noise correlation
\bea
\label{eq:langevinA}
\v (A_t)_{ij} = - {{\partial S_{\rm eff}} \over {\partial (A_t)_{ji}}} \v \t + \v (\xi _t)_{ij}, ~~~~~~
\langle \v (\xi _t)_{ij} \v (\xi _t')_{kl} \rangle _\xi = 2\v \t \d _{tt'} \d _{il} \d _{jk},
\ena
describe the evolution of the matrices on the step of the unit stochastic time $\v \t$.
They generate the Langevin equation for a loop variable 
\bea
\label{eq:langevin}
\v \phi _t (n) &=& \v \t n \left[ g \phi _t (n-1) - \phi _t (n) + g \phi _t (n+1) 
 + \sum_{k =0}^{n-2} \phi _t ( k ) \phi_t (n - k - 2) \right. \nonumber \\
 & & + \sum_{k=1}^{\infty} \sum_{m=0}^{\infty} G^{(1)} (k,m;1) \phi _t (n+k-2) \phi _{t+1} (m)  \nonumber \\
 & & \left. + \sum_{k=1}^{\infty} \sum_{\ell =0}^{\infty} G^{(2)} (\ell ,k;1) \phi _t (n+k-2) \phi _{t-1} (\ell) \right] + \v \zeta _t (n),
\ena
where $G^{(1)} (k,m;1) \equiv kG^{(0)} (k,m,t=1) $ and $G^{(2)} (k,m;1) \equiv mG^{(0)} (k,m,t=1) $ are 1-step marked propagators with a mark on the entrance loop and the exit loop, respectively.
$\v \zeta _t (n) \equiv N^{-1-{n \over 2}} n {\rm tr} ( \v \xi _t A_t^{n-1} )$ is the constructive noise term which satisfies the correlation
\bea
\label{eq:noisecorrelation}
\langle \v \zeta _t (n) \v \zeta _{t'} (m) \rangle _\xi = 2 \v \t \d _{tt'} {1 \over N^2} nm \langle \phi _t (n+m-2) \rangle _\xi .
\ena
Any observable $O(\phi)$ composed of loop variables is deformed, under the stochastic time 1-step progress $\v \t$, following the variation of $\phi _t (n)$ with the Langevin equation (\ref{eq:langevin}) and the noise correlation (\ref{eq:noisecorrelation}).
The generator of this $\v \t$ evolution corresponds to the Fokker-Planck (F-P) Hamiltonian $H_{\rm FP}$,
\bea
\label{eq:FPHdef}
\langle \v O(\phi ) \rangle _\xi &=& \langle \sum_m \v \phi _t (m) {\partial \over \partial \phi _t (m) } O(\phi ) + {1 \over 2} \sum_{m,n} \v \phi _t (m) \v \phi _t (n) {\partial ^2 \over \partial \phi _t (m) \partial \phi _t (n)} O(\phi ) \rangle _\xi \nonumber \\
 & & +{\rm O}(\v \t^{3 \over 2}) \nonumber \\
 & \equiv & - \v \t \langle H_{\rm FP} O(\phi ) \rangle _{\xi}.
\ena
We interpret $\phi _t (n)$ and $\pi _t (n) \equiv {\partial \over \partial \phi _t (n) }$ as the operators for creation and annihilation of the loop with length $n$ at time $t$, respectively.
They fulfill the following commutation relation:
\bea
\label{eq:commutation}
[\pi _t (n), \phi _{t'} (m) ] = \d _{tt'} \d _{nm}.
\ena
Then the F-P Hamiltonian is expressed as
\bea
\label{eq:FP1}
H_{\rm FP} = -{1 \over N^2} \sum_t \sum_{n=1}^{\infty} n L_t (n-2) \pi _t (n),
\ena
where $L_t (n)$ is defined by
\bea
\label{eq:generator}
L_t (n) &=& -N^2 \Biggl[ g \phi _t (n+1) - \phi _t (n+2) + g \phi _t (n+3) \Biggr. \nonumber \\
& & +\sum_{k=1}^{\infty} \sum_{m=0}^{\infty} G^{(1)} (k,m;1) \phi _t (n+k) \phi _{t+1} (m) 
 + \sum_{k=1}^{\infty} \sum_{\ell =0}^{\infty} G^{(2)} (\ell ,k;1) \phi _t (n+k) \phi _{t-1} (\ell) \nonumber \\
& & \left. + \sum_{k=0}^n \phi _t (k) \phi _t (n-k) + {1 \over N^2} \sum_{k=1}^{\infty} k \phi _t (n+k) \pi _t (k) \right],
\ena
and it satisfies the Virasoro algebra
\bea
\label{eq:virasoro}
[L _t (n), L _{t'} (m) ] = (n-m)\d _{tt'} L _t (n+m).
\ena

In order to take the continuum limit we introduce the minimum length of this matrix model $a$.
The continuum limit is realized by taking $a$ to zero simultaneously with $N$ to infinity.
It is called the double scaling limit.
According to the CDT model, we set the continuum loop length as $ L \equiv an $ and time as $T \equiv at $\cite{AL}.
We also define the cosmological constant $\Lambda$ by ${1 \over 2}e^{-{1 \over 2}a^2 \Lambda} = g$.
Here we introduce two scaling parameters $D$ and $D_N$.
Since the commutation relation of the continuum field operators satisfy,
\bea
\label{eq:commutator}
[\Pi  (L;T), \Phi  (L';T') ] = \d (T-T') \d (L-L'),
\ena
the scaling of the loop field operators can be described as $\Phi  (L;T) \equiv a^{-{1 \over 2}D} \phi _t (n) $ and $\Pi (L;T) \equiv a^{{1 \over 2}D-2} \pi _t (n)$ by using $D$.
To keep the effect of the first term of the last line in eq.(\ref{eq:generator}), the splitting interaction, we fix the scaling of the infinitesimal stochastic time as
\bea
\label{eq:stochastic}
d \t \equiv a^{{1 \over 2}D-2} \v \t.
\ena
Hence the existence of the continuum stochastic time requires $D>4$.
The terms in the second line of eq.(\ref{eq:generator}) express the characteristic interaction of this model, which corresponds to the I-K type interaction.
Thus we maintain these terms by redefining the 1-step propagator as $\tilde{G}^{(1)}(L_1, L_2 ; a) \equiv a^{-1} G^{(1)} (k,m;1)$, which gives the expression using the continuum lengths.
We define the string coupling constant as $G_{\rm st} \equiv {1 \over N^2} a^{D_N}$ with $D_N$.
Then we obtain the continuum limit of the F-P Hamiltonian ${\cal H}_{\rm FP}$ by $\v \t H_{\rm FP} \equiv d \t {\cal H}_{\rm FP}$, and it is written as
\bea
\label{eq:FPhamiltonian}
{\cal H}_{\rm FP}  & = & \int dT \int_0^{\infty} dL L \left[ a^{-{1 \over 2}D+3} {1 \over 2} \left( {{\partial ^2} \over {\partial L ^2}} - \Lambda \right) \Phi (L;T)  \right. \nonumber \\
 & & +  \int _0^{\infty} dL_1 \int _0^{\infty} dL_2 \tilde G ^{(1)} (L_1, L_2; a) \Phi (L+L_1;T) \Phi (L_2; T+a)  \nonumber \\
 & & +  \int _0^{\infty} dL_1 \int _0^{\infty} dL_2 \tilde G ^{(2)} (L_2, L_1; a) \Phi (L+L_1;T) \Phi (L_2; T-a)  \nonumber \\
 & & + \int _0^L dL_1 \Phi (L_1 ; T) \Phi (L - L_1 ;T) \nonumber \\
 & & \left.  + a^{-D_N -D+1} G_{\rm st} \int _0^{\infty} dL_1 L_1 \Phi (L+L_1; T) \Pi (L_1; T) \right]  \Pi(L ; T).
\ena
The first term on the r.h.s is the potential term, which means the propagation of a loop in an equi-temporal slice.
We have to remember that any propagation of the loop in one equi-temporal slice is not contained in the GCDT model.
Hence we expect this term to scale out in the continuum limit.
This fact requires $D<6$.
It should be noted that, thanks to the scaling of the cosmological constant, after summing up the first three terms in eq.(\ref{eq:generator}) the scaling of the propagation terms are enhanced two order higher compared with that of the original terms.
This enhancement makes the above restriction $D<6$ for $D$ consistent with another restriction $D>4$ from eq.(\ref{eq:stochastic}).

\begin{figure}[t]
\begin{center}
\includegraphics [width=100mm, height=25mm]{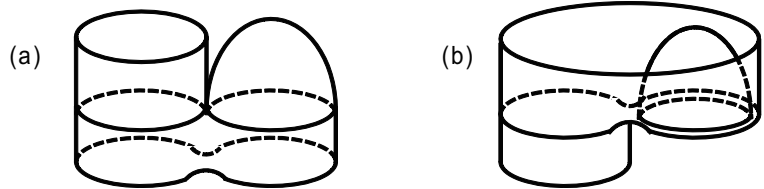}
\caption{(a) Splitting interaction and (b) I-K type interaction. The entrance loop propagates upward to the exit loop as the mother universe, while the baby universe must disappear to the vacuum before long. On the surface of propagating world-sheet, many baby universes of two types are attached as quantum effect.}
\label{fig:CDT3}
\end{center}
\end{figure}

The next two terms are I-K type interactions which are similar terms appeared in the non-critical string field theory model (Fig.\ref{fig:CDT3}(b))\cite{IK}.
The second and the third terms cause the annihilation of a loop with the length $L$ and creation of a loop with the length $L+L_1$ at the same time $T$.
They also create another loop with the length $L_2$ at the infinitesimal future time $T+a$ and infinitesimal past time $T-a$, respectively.
Then the lengths $L_1$ and $L_2$ are connected by the infinitesimal 1-step marked propagator.
The fourth term is the splitting interaction, which annihilates a loop with length $L$ and create two loops with the sum of their lengths $L$, simultaneously.
The last term expresses the merging interaction, which annihilates two loops and create one loop whose length is equal to the sum of the two annihilated loops.

While the splitting interaction is permitted in the GCDT model, the merging interaction should be forbidden because of the "causality" in a broad sense.
For this purpose we restrict the scaling parameter $D_N$ to satisfy
\bea
\label{eq:DN}
 D_N < -D+1.
\ena
Combining this condition and $4<D<6$, the value of $D_N$ can be fixed.
If $D_N<-5$ the causality is respected, while if $D_N>-3$ it is violated.
When $-5 \le D_N \le -3$, the condition of eq.(\ref{eq:DN}) must be taken into account.

The interpretation of the stochastic process is following.
If the stochastic process is not included, we have the exact CDT model.
Each stochastic time step $\v \t$ produces one baby universe in either of two ways, that is the ordinary splitting interaction and the I-K type interaction (Fig.\ref{fig:CDT3}).
While infinitesimal time or geodesic distance scales as $dT = a$, infinitesimal stochastic time scales by eq.(\ref{eq:stochastic}) with $4<D<6$.
This may be expressed as $d\t = a^{D_{\t}} \v \t $, with $0<D_{\t} <1 $.
This expression shows that $d \t$ goes to zero more slowly than $dT$.
From this we give a conjecture that the stochastic process, or the quantum effect occurs less frequently than the ordinary propagation in time.
Therefore we can expect the processes creating baby universes may not become dominant in the propagation.

It is interesting to suppose the change of the scaling parameter $D$.
The GCDT model is realized for $4<D<6$, and then the world-sheet might have reasonable number of baby universes of two types on its surface.
When $D>6$, the loop propagator contains too many stochastic processes, dominated by the propagation in the equi-temporal slice.
The string interactions are suppressed except for the merging interaction, whose probability depends on the scaling parameter $D_N$.
Hence, we can imagine that the loop propagator has too much length-fluctuation in the time evolution.
In $D=6$, each equi-temporal slice contains fertile stochastic processes including propagation, as well as splitting and I-K type interactions.
\footnote{In the S-D equation, as we will discuss later, an additional potential term concerning propagation $L ({\partial ^2 \over \partial L^2}-\Lambda ){W}(L)$, that is characteristic in the CDT model, survives in the scaling limit.}\
However, it is not desirable from the viewpoint of our model.

Finally, we derive the S-D equation from the continuum version of the Langevin equation.
The disc amplitude is the expectation value of a loop variable, which propagates and shrink to nothing eventually.
It is expressed in the continuum expression as
\begin{eqnarray*}
W(L) \equiv \langle \Phi (L;T) \rangle \equiv \int _0^{\infty}  dT \tilde{G}^{(1)} (L,0;T).
\end{eqnarray*}
With the help of eq.(\ref{eq:discsum}), at the level of the expectation value we can expect the relation 
\bea
\label{eq:disc}
\langle \int _0^{\infty} dL_2 \tilde{G}^{(1)}(L_1,L_2;a) \Phi (L_2; T+a) \rangle = W(L_1).
\ena
From this the S-D equation is derived as follows,
\bea
\label{eq:SD}
\int _0^L dL_1 W(L_1) W(L-L_1) +2 \int _0^{\infty} dL_1 W(L+L_1)W(L_1) =0. 
\ena
In terms of the Laplace-transformed variable $\tilde{W}(z) \equiv \int _0^{\infty} dL e^{-Lz} W(L) $, eq.(\ref{eq:SD}) is transformed to the expression in the Laplace space.
With the exchange of $z \rightarrow -z$, we obtain another equation for $\tilde{W}(-z)$.
Then using above two equations, we obtain the S-D equation as
\bea
\label{eq:LSD}
\tilde{W}(z)^2 +2 \tilde{W}(z) \tilde{W}(-z) +\tilde{W}(-z)^2 = \mu ,
\ena
where $\mu$ is some constant.
This S-D equation expresses the same form as the one of the non-critical closed string field theory except for the coefficient of the second term of the l.h.s..
It is replaced with a coefficient $2{\rm cos} \pi p_0 $, where the background momentum $p_0 ={1 \over m}$  corresponts to the central charge $c=1-{6 \over m(m+1)}$.
The last term of the effective action eq.(\ref{eq:actioneff}) is rewritten as $\sum_{t,t'}  C_{tt'} \sum_{k,m}^{\infty} G^{(0)}(k,m;1) \phi _t (k) \phi _{t'}(m) $ with an adjacency matrix $C_{tt'} \equiv \d _{t',t+1} + \d _{t',t-1} $.
If we adopt a twisted adjacency matrix $C^{(p_0)}_{tt'}= e^{i\pi p_0}\d _{t',t+1} + e^{-i\pi p_0} \d _{t',t-1}  $ as ref.\cite{Kos}, instead of our choice $C_{tt'} $, we can express the loop bilinear term of the effective action eq.(\ref{eq:actioneff}) as
\begin{eqnarray*}
 2 {\rm cos}\pi p_0 \sum _{t} \sum_{k=0}^{\infty} \sum_{m=0}^{\infty} G^{(0)}(k,m;1) \phi _t (k) \phi _{t+1}(m).
\end{eqnarray*}
We obtain the S-D equation of the non-critical string field theory exactly which has the right coefficient for the second term on the l.h.s. of eq.(\ref{eq:LSD}).

In conclusion, we have proposed the matrix model formulation to construct the 2D GCDT model.
Using the stochastic quantization approach and taking a continuum limit, we obtain the non-critical string field theory.
The scaling parameter $D$ and $D_N$ on the double scaling limit are fixed as eq.(\ref{eq:DN}) with $4<D<6$ by requiring the consistency to the CDT model.
We present one comment about a difference between the GCDT model and the discrete version of loop field theory formulated by our matrix model.
In our effective action eq.(\ref{eq:actioneff}) the sum of the 1-step propagation terms contains the propagation whose length of the exit loop or the entrance loop equal zero.
It is not contained in the original CDT model\cite{AL}.
We can cancel the $G^{(0)}(k,0;1)$ and $G^{(0)}(0,m;1)$ terms in the effective action by adding two fermionic degrees of freedom to the matrix model action.
It is done by assigning the Grassmann variables $C_t^{\a} , \overline{C}_t^{\a} (\a =1,2)$ to the link matrix connecting the sites in the same time slices, or $M_{tt} = A_t, C_t^{\a}, \overline{C}_t^{\a}$.
\footnote{The fermionic matrices $C_t^{\a}, \overline{C}_t^{\a}$ satisfy $(\overline{C}_t^{\a})_{ij}^* = - (C_t^{\a})_{ji}$.}\
Then the fundamental action is modified as $S_1 \equiv S - {\rm tr} \sum _t \overline{C}_t C_t + {g \over \sqrt{N}}{\rm tr} \sum _t \overline{C}_t A_t C_t $, where $S$ is the action of eq.(\ref{eq:actionAB}).
When we integrate out the fermionic degrees of freedom in $Z = \int {\cal D} A {\cal D} B {\cal D} B^{\dagger} {\cal D} C {\cal D} {\overline C} e^{-S_1 } $, the effective action has the additional term, $-N {\rm tr} \sum _t \sum _{n=1}^{\infty} {1 \over n} ( {g \over \sqrt{N}} )^n A_t^n$, which cancels the terms concerning to the loop-to-vacuum 1-step processes.
Therefore, we need not worry about whether the summation of the loop length begins from $n=0$ or $n=1$.

\section*{Acknowledgments}
The author would like to thank D. Suematsu for valuable comments.




\begin{thebibliography}{99}
%
\bibitem{DS} E. Br\'{e}zin and V. Kazakov, \PL{B236}(1990) 144; \\
M. Douglas and S. Shenker, \NP{B335}(1990) 635; \\
D. Gross and A. Migdal, \PRL{64}(1990) 127; \NP{B340}(1990) 333.
%
\bibitem{IK} N. Ishibashi, H. Kawai, \PL{B314}(1993)190, hep-th/9307045; \\
N. Ishibashi, H. Kawai, \PL{B322}(1994)67, hep-th/9312047.
%
\bibitem{JR} A. Jevicki, J. Rodrigues, \NP{B421}(1994)278, hep-th/9312118.
%
\bibitem{AJ} J. Avan, A. Jevicki, \NP{B469}(1996) 287, hep-th/9512147.
%
\bibitem{Mog} T. Mogami, \PL{B351}(1995) 439, hep-th/9412212.
%
\bibitem{KK} I. Kostov, \NP{B376}(1992) 539, hep-th/9112059; \\
V. A. Kazakov, I. Kostov, \NP{B386}(1992) 520, hep-th/9205059.
%
\bibitem{Kos} I. Kostov, \PL{B344}(1995) 135, hep-th/9410164; \\
I. Kostov, \PL{B349}(1995) 284, hep-th/9501135.
%
\bibitem{EKN} D. Ennyu, H. Kawabe, N. Nakazawa, \PL{B454}(1999) 43, hep-th/9902001.
%
\bibitem{Nak} N. Nakazawa, \MPL{A10}(1995) 2175, hep-th/9411232; \\
N. Nakazawa, D. Ennyu, \PL{B417}(1998) 247, hep-th/9708033.
%
\bibitem{AL} J. Ambj{\o}rn, R. Loll, \NP{B536}(1999) 407, hep-th/9805108.
%
\bibitem{ALWZ} J. Ambj{\o}rn, R. Loll, W. Westra, S. Zohren, \JHEP{0712}(2007) 017, \\
arXiv:0709.2784[gr-qc].
\bibitem{ALWWZ1} J. Ambj{\o}rn, R. Loll, Y. Watabiki, W. Westra, S. Zohren, \JHEP{0805}(2008) 032, arXiv:0802.0719[hep-th].
%
\bibitem{ALWWZ2} J. Ambj{\o}rn, R. Loll, Y. Watabiki, W. Westra, S. Zohren, \PL{B665}(2008) 252, arXiv:0804.0252[hep-th].
%
\bibitem{FSW} H. Fuji, Y. Sato, Y. Watabiki, \PL{B704}(2011) 582, arXiv:1108.0552[hep-th].
%
\end{thebibliography}
\end{document}